\documentclass[conference,onecolumn,draftcls]{IEEEtran}
\IEEEoverridecommandlockouts

\makeatletter
\def\markboth#1#2{\def\leftmark{\@IEEEcompsoconly{\sffamily}\MakeUppercase{\protect#1}}%
\def\rightmark{\@IEEEcompsoconly{\sffamily}\MakeUppercase{\protect#2}}}
\makeatother

\usepackage[utf8x]{inputenc}
\usepackage[english]{babel} 
\usepackage{ucs} 
\usepackage{amsmath} 
\usepackage{amsfonts} 
\usepackage{amssymb} 
\usepackage{amsthm}
\usepackage{array}
\usepackage{verbatim}
\usepackage{listings}
\usepackage{stfloats}

\usepackage{algorithm} 
\usepackage{algorithmic} 
\usepackage{url} 
  \usepackage{enumerate}
  \usepackage[table]{xcolor}

\ifCLASSINFOpdf
  \usepackage[pdftex]{graphicx}
  \usepackage{subfigure}
  \usepackage{epstopdf}
   \graphicspath{{./img/}}
   \DeclareGraphicsExtensions{.eps,.ps}
\else
  \usepackage[dvipdf]{graphicx}
  \usepackage{subfigure}
  \graphicspath{{./}}
   \DeclareGraphicsExtensions{.eps,.ps}
\fi

\newcommand{\x}{\mathbf{x}}

\newcommand{\Ex}[2]{{\textnormal{E}_{#1}\left[#2\right]}}

\title{
Improving Third-Party Relaying for LTE-A: A Realistic Simulation Approach 
}

\author{
  \IEEEauthorblockN{Felipe G\'omez-Cuba,~\IEEEmembership{Student Member,~IEEE,} Francisco J. Gonz\'alez-Casta\~no}
    \IEEEauthorblockA{Information Technologies Group, AtlantTIC, University of Vigo, Spain\\  fgomez@gti.uvigo.es, javier@det.uvigo.es}
}
\begin{document} 
\maketitle
\markboth{DRAFT}{DRAFT}
\begin{abstract}
In this article we propose solutions to diverse conflicts that result from the deployment of the (still immature) relay node (RN) technology in LTE-A networks. These conflicts and their possible solutions have been observed by implementing standard-compliant relay functionalities on the Vienna simulator.

As an original experimental approach, we model realistic RN operation, taking into account that transmitters are not active all the time due to half-duplex RN operation. We have rearranged existing elements in the simulator in a manner that emulates RN behavior, rather than implementing a standalone brand-new component for the simulator. We also study analytically some of the issues observed in the interaction between the network and the RNs, to draw conclusions beyond simulation observation.

The main observations of this paper are that: $i$) Additional time-varying interference management steps are needed, because the LTE-A standard employs a fixed time division between eNB-RN and RN-UE transmissions (typical relay capacity or throughput research models balance them optimally, which is unrealistic nowadays); $ii$) There is a trade-off between the time-division constraints of relaying and multi-user diversity; the stricter the constraints on relay scheduling are, the less flexibility schedulers have to exploit channel variation; and $iii$) Thee standard contains a variety of parameters for relaying configuration, but not all cases of interest are covered. 
% iv) Even though there is literature considering either planned optimal location of relays or random locations; we introduce an intermediate approach based on admission control to improve performance on the random-location approach while still avoiding the need for planning.
\end{abstract}


\section{Introduction}
\label{sec:introduction}
3GPP LTE-A relies on several improvements such as carrier aggregation (CA), Multiuser MIMO (MU-MIMO) and femtocells to meet the IMT-Advanced requirements for 4G standards \cite{mogensen2009lte}. One particularly interesting aspect is relaying \cite{journals/cm/LoaWSYCHX10}.

LTE-A relays behave like any other Evolved Node B (eNB) from the point of view of User Equipment (UE) \cite{LTE36216PHYrelay}. The eNB of the cell where the RN is located, called the Donor eNB (DeNB), must be aware of the relay and provide proxy functionality to its interfaces with the rest of the network infrastructure. This proxying, as well as UE traffic forwarding, is carried out by a DeNB-RN LTE-A connection that requires the RN to behave temporarily as a UE from the point of view of the DeNB. Hereafter, we will call this link the \textbf{relay} link; \textbf{access} links are those between UE and RN and \textbf{direct} links, those between UE and DeNBs.

LTE-A networks have a series of high-performance features that include intelligent UE scheduling, as a function of instantaneous channel states; multi-antenna techniques; interference management; and admission control \cite{mogensen2009lte,Specification2012c}. As shown in this paper, relays may alter these features. In the literature, RN location optimization, which may be constrained to mitigate conflicts, is a common approach, but random location of RN and femtocells is sometimes considered \cite{Karaer2009optimization,conf/vtc/SalehRHR09,conf/vtc/SalehRHRRW09Performance,journals/jece/SalehRHR10}.

In this article we propose solutions to diverse conflicts that result from the deployment of the still immature relay node (RN) technology in an LTE-A network. These conflicts have been identified by implementing standard-compliant relay functionalities on top of the well-known Vienna LTE System Level Simulator \cite{vienaVTC2010}. 
% As an original experimental approach, we model realistic RN operation, taking into account that transmitters are not active all the time due to half duplex RN operation. 
Implementing RN firmware is too complex, but, without loss of realism, we have rearranged existing elements  (eNB, UE, scheduler) in a manner that -seen as a black box- emulates the behavior of a relay, rather than implementing a standalone brand-new component for the simulator. 
We also study analytically some of the issues observed in the interaction between the network and the RNs, to draw conclusions beyond simulation observation.

As far as we know, the conflicts we have identified related to interference management, scheduling and half-duplex relay operation have not been reported so far. Their characterization and possible solutions are the main contributions of this paper.

% iv) Even though there is literature considering either planned optimal location of relays or random locations; we introduce an intermediate approach based on admission control to improve performance on the random-location approach while still avoiding the need for planning. 

The rest of this paper is organized as follows: Section \ref{sec:standard} describes relaying in LTE-A Release 10. Section \ref{sec:scenario} describes our model and the configuration of the Vienna simulator to implement it. Section \ref{sec:interference} describes relaying problems with interference and our proposal to minimize their effects. Section \ref{sec:schedulers} describes how scheduling is hampered by relaying time constraints, even when measures are applied to improve it. We study this problem analytically and discuss the effect of all relevant parameters involved. Section \ref{secVI} proposes a possible approach to mitigate this problem.
% Section \ref{sec:AC} describes our admission control algorithm. 
Finally, section \ref{sec:conclusions} concludes the paper.

\section{Brief Introduction to Relaying in the LTE-A Standard}
\label{sec:standard}

LTE-A relays behave like any other eNB \cite{LTE36216PHYrelay}. The DeNB is aware of relay presence and provides proxy functionality to its \texttt{S1} and \texttt{X2} interfaces towards the rest of the E-UTRAN. The RN is a Layer-3 device featuring two physical interfaces, with UE and eNB functionalities in the relay link and access link, respectively \cite{LTE36216PHYrelay,LTE36213PHYproc}. Information must be forwarded using tunneling, according to \cite{conf/globecom/HuangUAHB10}, where LTE-A relays behave like L3 bridges, and the RN queues IP packets independently and transfers them between the User Equipment (UE) and the Donnor-eNB (DeNB). RNs become attached to a DeNB using the \texttt{relay\_attachment\_procedure} \cite{LTE36213PHYproc}. When the RN is in receive mode, it marks frames as MBMS to make the UEs oblivious to that mode.

\begin{figure}[!ht]
\centering
 \includegraphics[width=.7\columnwidth]{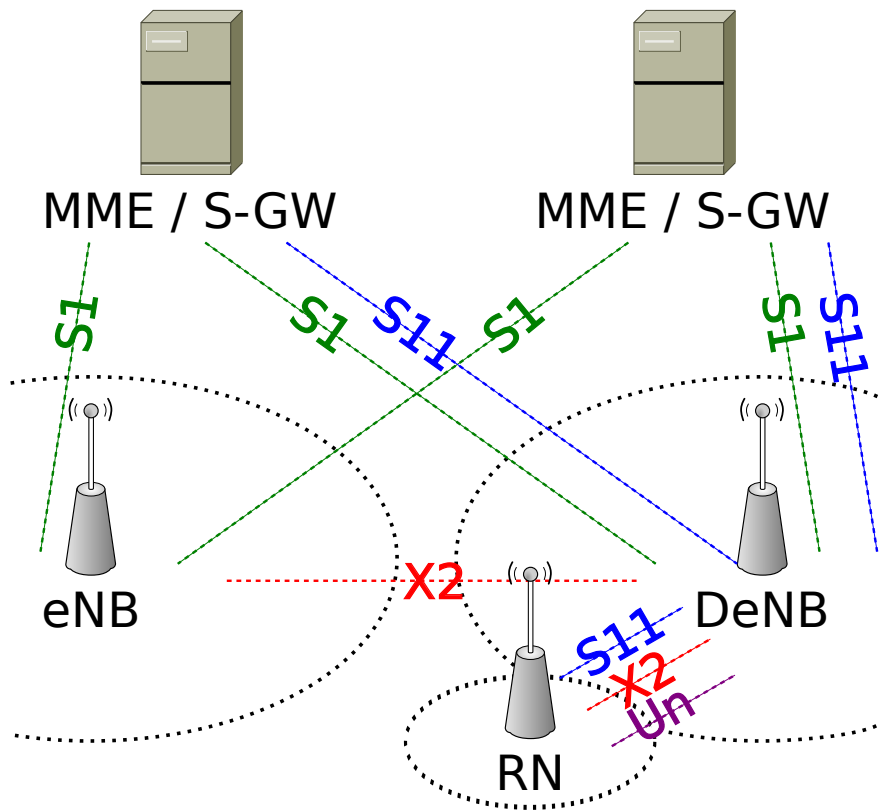}
 \caption{Relay architecture in LTE-A R10 (from figure 4.7.2-1 of \cite{LTE36300UTRANgen})}
 \label{fig:relay}
\end{figure}

% \begin{table*}[!ht]
% \centering
% \caption{LTE-A TDD number of subframes dedicated to the relay link and to the UEs.}
% \label{tab:tdd}
% % \rowcolors{1}{white}{lightgray}
% \begin{tabular}{l|cccccccccccccccccccc}
% \texttt{SubframeConfigurationTDD} & 0& 1 & 2& 3& 4& 5& 6& 7 & 8& 9 & 10 & 11& 12& 13& 14& 15& 16 & 17 & 18\\\hline
% Uplink-downlink setup & 1& 1& 1& 1& 1& 2& 2& 2 & 2& 2 & 2 & 3 & 3 & 4& 4& 4& 4 & 4 & 6\\\hline
% D subframes& 4&4&4&4&4& 6&6&6&6&6&6&6&6& 7&7&7&7&7& 3	\\
% D RN subframes&	1&1& 2&2&2& 1&1& 2&2& 3&3& 2&3& 1&2&2&3&4& 1 \\\hline
% U subframes& 4&4&4&4&4& 2&2&2&2&2&2&3&3& 2&2&2&2&2& 5	\\
% U RN subframes 1&1&1&1& 2& 1&1&1&1& 1&1&1&1& 1&1&1&1& 1&1&1 \\
% \end{tabular}
% % \begin{tabular}{m{2cm}m{2cm}|m{2cm}m{2cm}m{2cm}m{2cm}} 
% % TDDD subframe configuration&	Uplink-downlink setup&	D subframes&	D RN subframes&	U subframes&	U RN subframes\\\hline
% % 0&	1&	4&	1&	4&	1\\
% % 1&	1&	4&	1&	4&	1\\
% % 2&	1&	4&	2&	4&	1\\
% % 3&	1&	4&	2&	4&	1\\
% % 4&	1&	4&	2&	4&	2\\
% % 5&	2&	6&	1&	2&	1\\
% % 6&	2&	6&	1&	2&	1\\
% % 7&	2&	6&	2&	2&	1\\
% % 8&	2&	6&	2&	2&	1\\
% % 9&	2&	6&	3&	2&	1\\
% % 10&	2&	6&	3&	2&	1\\
% % 11&	3&	6&	2&	3&	1\\
% % 12&	3&	6&	3&	3&	1\\
% % 13&	4&	7&	1&	2&	1\\
% % 14&	4&	7&	2&	2&	1\\
% % 15&	4&	7&	2&	2&	1\\
% % 16&	4&	7&	3&	2&	1\\
% % 17&	4&	7&	4&	2&	1\\
% % 18&	6&	3&	1&	5&	1
% % \end{tabular}
% \end{table*}

% \begin{figure}[!hb]
% \centering
%  \includegraphics[width=.85\columnwidth]{LTE_relay_subframe}
%  \caption{LTE relay subframe in TTD setup \#1}
%  \label{fig:subframe}
% \end{figure}

In some subframes of the 5-ms LTE frame, rather than serve its UE (access link), the RN is served by the DeNB (relay link). This depends on the configuration parameters \texttt{SubframeConfigurationFDD} and \texttt{SubframeConfigurationTDD}, for  the different multiplexing modes. The former is used to choose between eight subframe partitions of relay and access links: $1/7,2/6,,3/5\dots6/2,7/1$. The latter allows the TDD setups in table 5.2-2 of \cite{LTE36216PHYrelay}.

LTE relay performance has been assessed by several authors \cite{conf/vtc/SalehRHR09,conf/vtc/SalehRHRRW09Performance,journals/jece/SalehRHR10,journals/ejwcn/DopplerRWRW09,conf/icc/RongEH10}. 
For example, Saleh et al \cite{conf/vtc/SalehRHR09}\cite{conf/vtc/SalehRHRRW09Performance} studied RN and Pico-eNB throughput gains for the worst 10th percentile of LTE-A network users, with a simulation setup very close to reality. They considered a hexagonal lattice with three 120\textdegree sectors per eNB, helped by 5 to 12 small cells. From the resulting  iso-performance curves (number of eNB per km$^2$ vs. number of small cells per km$^2$ with the same performance), they concluded that the relays must be cheaper  than $~1/30$ of the cost of the eNBs for the approach to be more advantageous than a mere increment in eNB density.

Even though some studies have dealt with Amplify-and-Forward relaying \cite{conf/icc/RongEH10}, non-orthogonal Decode and Forward relaying has been considered more frequently. Within this system, the DeNB transmits to its direct UE in the RN-UE phase. We also consider this system, and,  from the three 3GPP relay types in \cite{journals/jece/SalehRHR10}, \textit{Type 1} inband relays, without antenna isolation between relay and access links.

\section{Implementation of the Relaying Scenario}
\label{sec:scenario}

The Vienna LTE Downlink System Level Simulator evaluates common MAC and RRC tasks during a series of LTE TDD subframes (referred as TTIs). Scheduling granularity is defined by \textbf{resource blocks} (RBs), whereas traffic delivery/error measurement granularity is defined by \textbf{transport blocks} \cite{136211LTEPHYr10}. The simulator implements different objects for schedulers, eNBs, UEs, path-loss maps of terrain, etc. Unlike Femtocell simulation,  relay simulation is still unsupported.

The schema we have developed is aimed at replicating at system level the behavior of LTE-A networks with relays, in order to study how interference, scheduling and data delivery are affected by the addition of these relays, with the realistic (novel) approach that not all transmitters are active all the time due to half duplex RN operation. As previously stated, we avoided implementing complex internal characteristics of RN firmware by rearranging existing elements of the simulator (eNB, UE, scheduler), in a manner that -seen as a black box- emulates the behavior of a relay. The alternative would be to implement a standalone brand-new component but this would duplicate many lines of code already included in eNBs or UEs anyway (for transmission and reception purposes respectively).

\begin{figure}[!ht]
 \centering
 \includegraphics[width=.75\columnwidth]{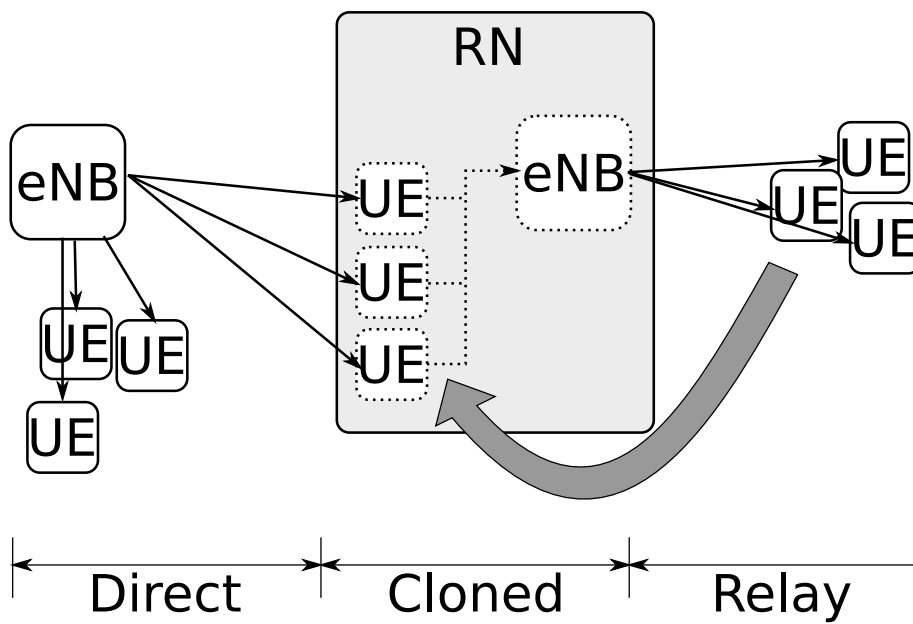}
 \caption{Design of a setup for the simulator that emulates the behavior of an RN using only existing components. }
 \label{fig:RNdesign}
\end{figure}

Figure \ref{fig:RNdesign} illustrates the setup. For each RN, a virtual eNB is created. Normal UEs are created and attached to the eNB using default simulator procedures (which include virtual eNBs). Finally, each RN creates a clone of each of its UEs and attaches them to its serving DeNB. DeNB downlink traffic is scheduled for direct and cloned UEs in the subframes when the RN is supposed to be receiving, and the virtual eNB transmits empty frames. Cloned UEs forward the data received to the outbound traffic buffer of their associated virtual eNB. In RN transmission subframes, the DeNB only schedules direct users and the virtual eNB schedules relay UEs, which receive the data that their clones forward. Thus, outside the grey box in figure \ref{fig:RNdesign} the network behaves exactly as if a true RN was implemented, and no modification of simulator core elements is required. The novelty of our implementation lies in the cloned UEs and their traffic-forwarding behavior, and in restricting the UEs that the 
schedulers are allowed to serve at certain given times. The virtual eNB at the relay follows the original femtocell configuration of the official simulator distribution.

\begin{figure}[!ht]
\centering
\subfigure[Without RN]{
 \includegraphics[width=.75\columnwidth]{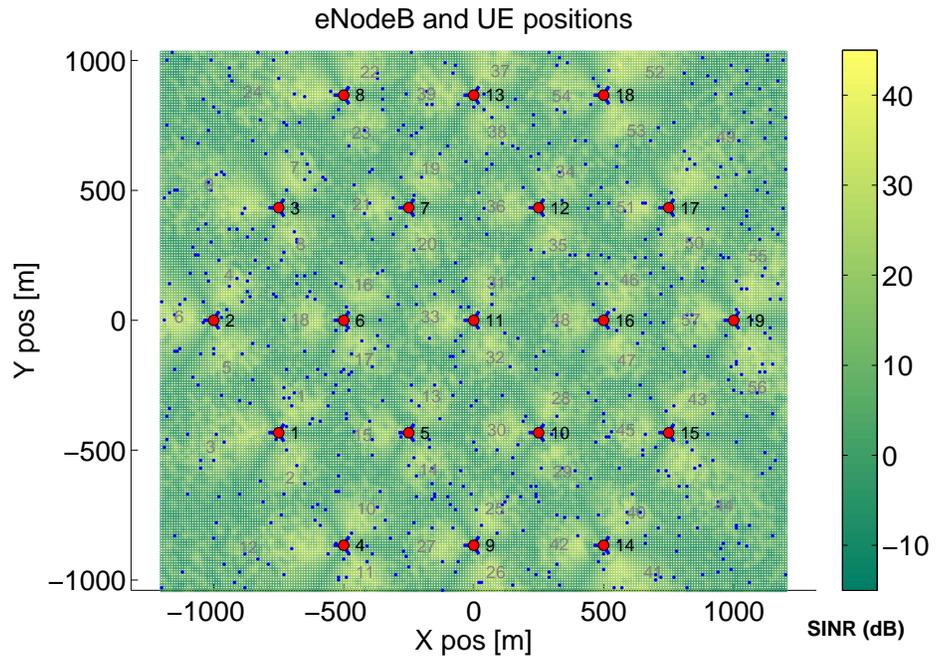}
 \label{fig:mapNO}
 }
\subfigure[With RN]{
 \includegraphics[width=.75\columnwidth]{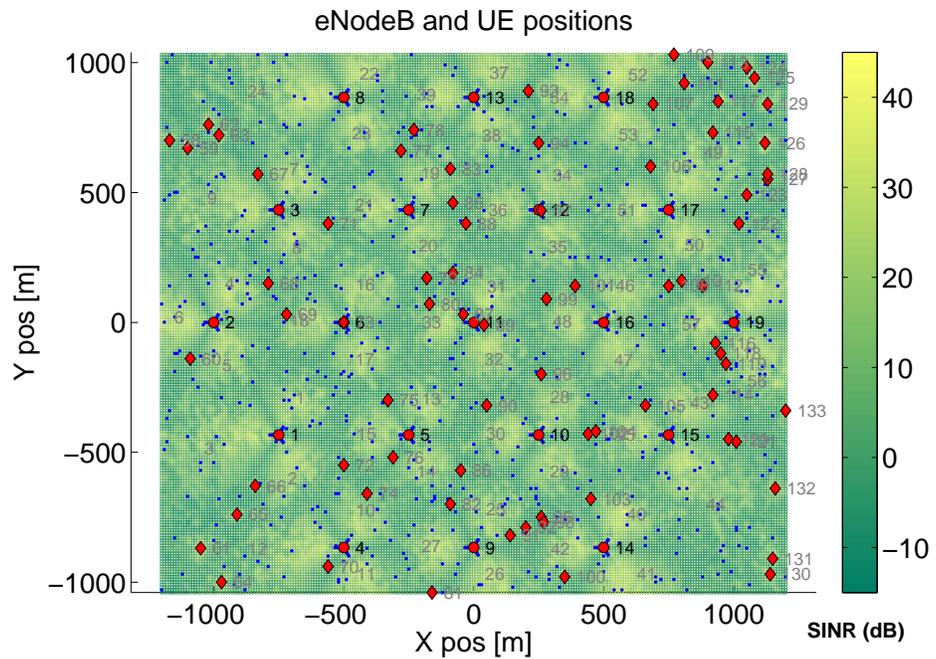}
 \label{fig:mapRN}
 }
 \caption{Comparison of simulation setups and SINR with and without RNs}
 \label{fig:map}
\end{figure}

Figure \ref{fig:map} shows the insertion of 200 randomly located RNs in the simulation. The valleys in subfigure \ref{fig:mapRN} are darker than those in subfigure \ref{fig:mapNO}, representing worse Signal to Interference and Noise Ratio (SINR).

\section{Mitigating Interference in Half-Duplex Relay Operation}
\label{sec:interference}

Typically, a $1/2$ relaying ratio is not optimal. For instance, if relay links are better than access links, it is beneficial to balance link traffic by allocating more time to the weakest ones, so that the RN does not have to drop packets correctly delivered by the DeNB, thus wasting resources. When the spectral efficiencies of relay and access links are known ($\rho_r,\rho_a$), optimal end-to-end spectral efficiency is as follows \cite{conf/vtc/SalehRHRRW09Performance,journals/jece/SalehRHR10}:
% \begin{equation}
% \label{eq:alphaopt}
% \alpha_{opt}=\frac{\rho_a}{\rho_r+\rho_a} 
% \end{equation}
% so that the end-to-end throughput is
\begin{equation}
\label{eq:paralell}
\theta_r=\frac{\rho_r\rho_a}{\rho_r+\rho_a} 
\end{equation}

However, the standard defines a fixed configuration that is applied to the whole system.
% (table \ref{tab:tdd})
This implies that throughput balance is sacrificed to achieve a common timing across the network. This is bad from a relaying point of view, as it necessarily lowers performance.

When the network enters its \textit{relaying phase} (subframes marked for RN operation), access links are free of transmissions. From a radio perspective, at these moments the network should experience the same interference levels as if there were no relays at all (fig \ref{fig:mapNO}). Next, when the network enters its \textit{access phase} (subframes for the RNs to serve their users), DeNBs do not schedule relay links, and the RNs transmit to their UEs. From a radio perspective, this second phase necessarily has higher interfering power due to the higher density of transmitting entities (fig \ref{fig:mapRN}).

The key question when introducing RNs in an LTE-A network is whether the enhanced reachability of ill-located UE is worth the greater interference or not. However, since throughputs are imbalanced, access links do not necessarily have to transmit continuously during the whole access phase; they are likely to have less buffered data than the capacity of all the RBs available. Thus, a crude form of interference mitigation can be implemented by reducing the number of subframes that RNs actively use. It is possible to grant some bonus subframes to direct users in the access phase. In this case there is no RN transmission, and interference is still the same as that in an RN-less network. The price RN schedulers pay is fewer subframes to choose from.
%, Proportional-Fair (PF) schedulers in RN achieve less multi-user gain. 
We cover scheduling conflicts in section \ref{sec:schedulers}.

In our schema, there are three types of subframes:
\begin{itemize}
 \item \textit{b-subframes:} All flows connected to the DeNB are scheduled. This is the relaying phase and \textbf{must} be planned in compliance with the standard.
 \item \textit{u-subframes:} All the UEs, but not RN flows, are scheduled. This is the normal behavior of the access phase in the standard. 
 \item \textit{d-subframes:} This is our proposal. These subframes belong to the access phase but only direct UEs are scheduled in them. They are neither defined in the standard nor incompatible with it. We propose combining u-subframes and b-subframes in search of good operation regimes.
\end{itemize}

The standard mandates a strict number of b-subframes for the whole network, creating a throughput imbalance. When there are more RBs in u-subframes than needed, some are left empty, but other users in the surroundings are oblivious to this fact and will still consider those RBs to be subject to RN-level interference. In practice, RN interference increases with the number of u-subframes, even if the RNs have no data to deliver. Therefore, we recommend maintaining a balance between u- and d-subframes. For instance, for \texttt{subframeConfigurationTDD}=$6$, when only one out of six subframes is a b-subframe, it is wasteful to use five u-subframes. The simulation shows better performance for 2 u-subframes and 3 d-subframes. 

\begin{figure}[!ht]
\centering
\subfigure[5 u-subframes]{
 \includegraphics[width=.75\columnwidth]{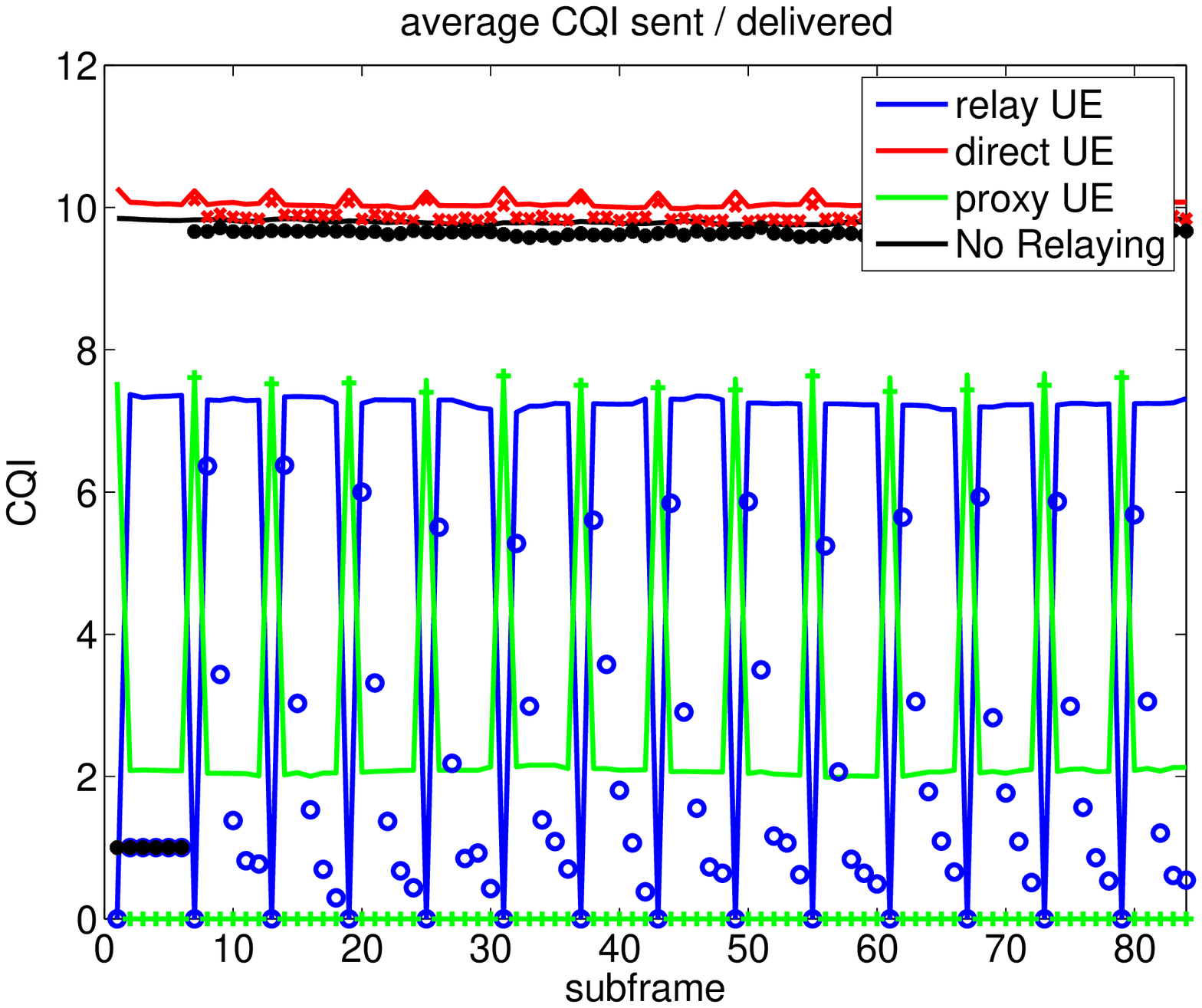}
 \label{fig:cqi5}
 }
\subfigure[3 d-subframes followed by 2 u-subframes]{
 \includegraphics[width=.75\columnwidth]{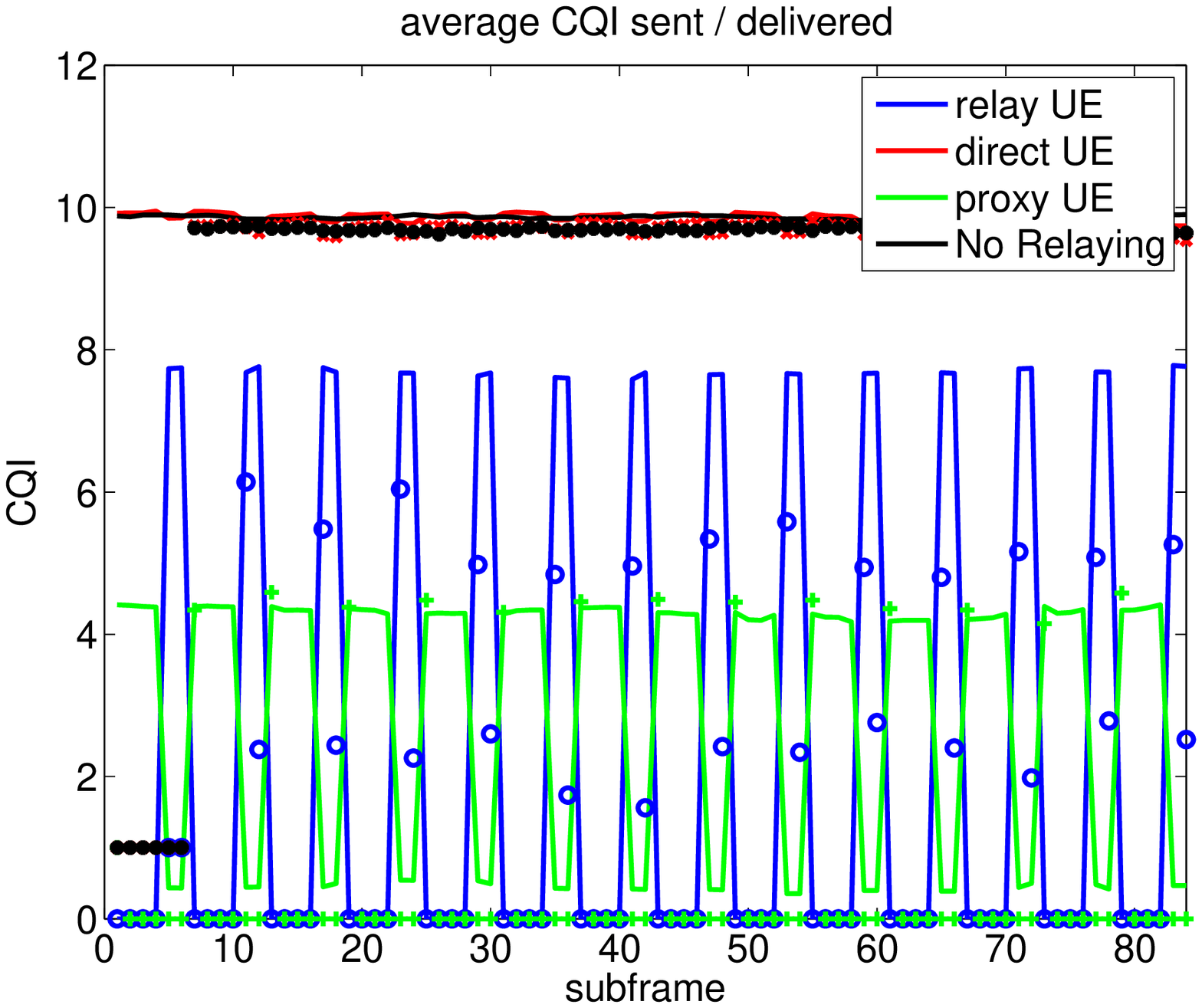}
 \label{fig:cqi2}
 }
 \caption{CQI in 84 simulation subframes for \texttt{SubframeConfigurationTDD}=$6$, showing RN scheduling for interference mitigation.}
 \label{fig:cqi}
\end{figure}

Figure \ref{fig:cqi} shows the Channel State Indicator (CQI) reported by the UE (solid lines) and the modulation and coding scheme (MCS) index as actually delivered by the eNBs (bullets). If RBs were fully occupied the two curves should be equal, as in the simulation without relays (black) and the direct and proxy UE with relays (red and green, respectively). 

In subfigure \ref{fig:cqi5} all subframes in the access phase are u-subframes. There is a periodical pattern of six subframes, starting with a relay phase with one b-subframe. In the access phase, relayed UEs are only scheduled in subframes 2 and 3, whereas in subframes 4 to 6 they have negligible traffic. However, direct UEs do have the same reported CQI in all the access subframes, which means that interference is avoidable.

In subfigure \ref{fig:cqi2} we repeated the simulation with one relay b-subframe followed by three d-subframes only with direct users, followed by two d-subframes with direct and relayed users. The MCSs of relayed users coincide, whereas the MCSs of direct users improve in three of the five access subframes. This example shows how interference can be mitigated, and there is additional margin to improve the gains with optimization methods.

\section{Scheduling Degradation in Relaying Scenarios}
\label{sec:schedulers}

\subsection{Description}
LTE-A allows for UE traffic to be scheduled in time and frequency slots called RBs. Scheduler implementation details are left to the vendors, so there are different commercial products with simple or optimized schedulers. Well-known scheduling algorithms featured in the Vienna simulator include Round-Robin (RR), Max Throughput, and Proportional-Fair Scheduling (PF).

During simulation, it was consistently observed that the performance improvements after adding relays appeared to be smaller with PF schedulers than with RR schedulers.
% (Fig \ref{fig:tputs}). 
On the one hand, RR can be considered a fair time multiplexing; therefore, the effects of relays on its gain is due to the changes in spectral efficiency. On the other hand, PF schedulers exploit multiuser diversity, and thus, with these, the typical user channel is better than its average fading realization. However, when a user is attached to an RN, relay link traffic must take place in specific subframes of the LTE radio interfaces regardless of how a PF scheduler would handle it. In other words, half-duplex operation overrides scheduling and causes a conflict between relaying and 
%the ability of t of relaying conflicts with 
PF scheduling, 
% which has less time slots or users to pick from, thus necessarily
hampering multi-user diversity gain.

% \begin{figure}[!ht]
% \centering
% \subfigure[Round Robin]{
%  \includegraphics[width=.85\columnwidth]{ModesThroughput-RR}
%  \label{fig:tputsRR}
%  }
% \subfigure[Proportional Fair]{
%  \includegraphics[width=.85\columnwidth]{ModesThroughput-PF}
%  \label{fig:tputsPF}
%  }
%  \caption{Average throughput per user over $N$ simulations, versus half-duplex factor (depending on \texttt{SubframeConfigurationTDD}), for the two most common schedulers, RR and PF.}
%  \label{fig:tputs}
% \end{figure}

\begin{figure}[!ht]
 \centering
 \includegraphics[width=.5\columnwidth]{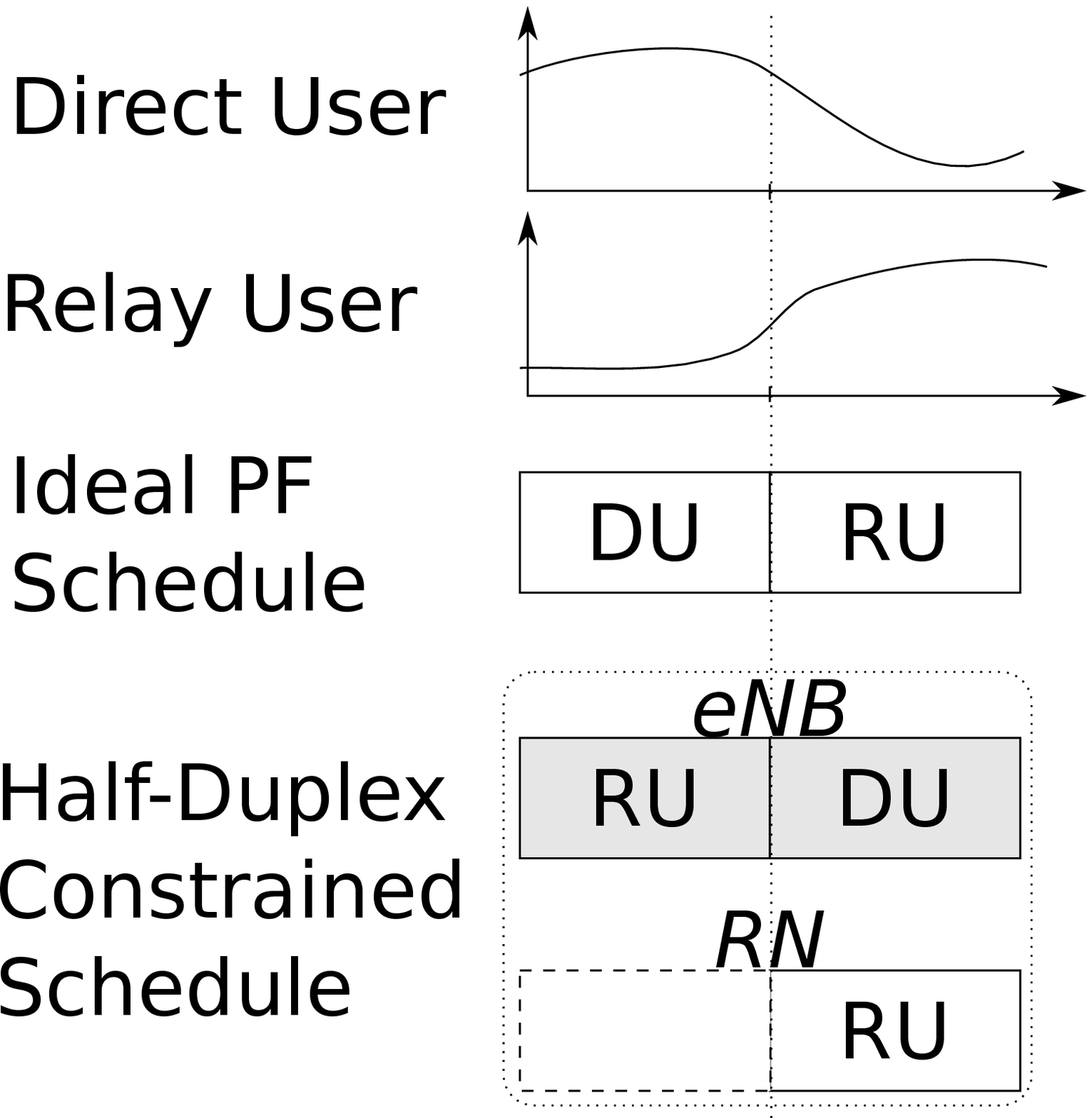}
 \caption{Example of PF scheduler conflict due to the half-duplex relaying constraint for two users: a direct user (DU) and a relay user (RU). }
 \label{fig:schedulerproblem}
\end{figure}

Figure \ref{fig:schedulerproblem} illustrates the conflict between PF scheduling and relaying. For the channel realizations of two users (rows 1 and 2), the ideal PF schedule is shown in row 3, whereas the necessary schedule to meet half-duplex relay operation is shown in row 4. Users are displaced from their ideal channel realizations to worse ones. Furthermore, direct users are moved to worse channels that \textit{also} have higher interference due to the simultaneous transmission of the relay.

We can study this effect analytically. The general convergence of PF schedulers was analyzed in \cite{Kushner2004}. For a single cell scheduler, the main result was that, for each user $i$, average throughputs $\overline\theta_i$ satisfy the first order Ordinary Differential Equation (ODE) system:
\begin{equation}
\label{eq:ode}
 \begin{array}{ccccc}
  \dot{\overline\theta}_1&=&\Ex{t:s[t]=1}{\theta_1[t]}&-&\overline\theta_1\\
  \dot{\overline\theta}_2&=&\Ex{t:s[t]=2}{\theta_2[t]}&-&\overline\theta_2\\
  \vdots&&\vdots&&\vdots\\
  \dot{\overline\theta}_n&=&\Ex{t:s[t]=n}{\theta_n[t]}&-&\overline\theta_n\\
 \end{array}
\end{equation}
where $s[t]\in 1\dots n = \arg_i\max \frac{\theta_i[t]}{\overline\theta_i}$ is the proportional fair scheduling function that selects the user with the highest ratio between its instantaneously achievable throughput and its average. 

% Assuming that each user experiences Rayleigh fading $h_i$ with parameter $\lambda_i$ and that capacity is, in the low-SNR regime, approximately linear with SNR,

Without relays, under Rayleigh fading with parameter $\lambda_i$, PF achieves $\overline{\theta}_{i,PF}=\frac{\sum_{j=1}^{n}\frac{1}{j}}{n}\frac{1}{\lambda_i}$ \cite{Kushner2004}, while RR achieves $\overline{\theta}_{i,RR}=\frac{1}{\lambda_i}$
% \begin{equation}
% \label{eq:PFratesimple}
%  \begin{split}
%   \overline{h}_i&=\Ex{h_i[t]/\overline{h}_i>h_j[t]/\overline{h}_j\forall j}{h_i[t]}\\
%     &=\frac{\sum_{j=1}^{n}\frac{1}{j}}{n}\frac{1}{\lambda_i},\\
%  \end{split}
% \end{equation}
% where the first term represents multiuser gain versus RR operation, $\overline\theta_{i,RR}=\frac{1}{\lambda_i}$. 
% The integral to compute the average is formulated in appendix \ref{app:schedulerintegral}.

\subsection{Scheduler Analysis with Relays}

Let us now consider a single-cell with a relay setup consisting of $\tau_r$ relay subframes, in which the DeNB can schedule direct and relay links, followed by $\tau_a$ access subframes in which the DeNB can only schedule direct links while the relays operate as eNBs and schedule their own users. Users $1$ to $n_d$ are direct users, and users $n_d+1$ to $n_d+n_r$ are relayed users. 

In the relay phase, allocating a direct transmission only achieves the benefit of that particular data delivery, whereas allocating a relay transmission achieves the additional benefit of ensuring that the relay will have data to deliver in the following access phase. Therefore, we consider an \textit{incentivized} PF scheduler as in \cite{Gitlin}, with an incentive parameter $\beta>1$. The instantaneous throughputs of the flows in the DeNB scheduler will then be:
\begin{equation}
 \theta_i[t]=\begin{cases}
              h_i&\{i\leq n_d\}\cup\{t_{\mod(\tau_r+\tau_a)}< \tau_r\}\\
              0&\{i> n_d\}\cap\{t_{\mod(\tau_r+\tau_a)}\geq \tau_r\}\\
             \end{cases}
\end{equation}

In the first fraction of the frame, with relative duration $\alpha=\frac{ \tau_r}{\tau_r+\tau_a}$, all users can be allocated, while in the second, with relative duration $1-\alpha$, only direct users can access the channel. Thus, taking the averages separately in these two fractions, we obtain the original PF scheduler in the first fraction, and a new PF realization for a subset of the users in the second fraction.
\begin{equation}
\label{eq:oderelay}
% \begin{split}
 \overline{\theta}_i=\begin{cases}
   \alpha\Ex{h_i[t]/\overline{\theta}_i>b_j h_j[t]/\overline{\theta}_j\forall j}{h_i[t]}&i\leq n_d\\
   \quad+(1-\alpha)\Ex{h_i[t]/\overline{\theta}_i>h_j[t]/\overline{\theta}_j\forall j\leq n_d}{h_i[t]}\\
   \alpha\Ex{h_i[t]/\overline{\theta}_i>b_j h_j[t]/\overline{\theta}_j\forall j}{h_i[t]}&i> n_d\\
           \end{cases}
% \end{split}
\end{equation}
where
\begin{equation}
 b_j=\begin{cases}
          1&j\leq n_d\\
	  \beta_j& n_d<j\\
         \end{cases}
\end{equation}

Consequently, at the end of the frame, direct flows will be more likely to have gained access to the channel than relay flows, an effect that can be compensated with incentive $\beta$. The resulting ODE system, formulated in Appendix \ref{app:schedulerintegral}, has two sets of equations: direct flow throughputs receive two contributions, whereas relay throughputs receive only one contribution. These three contributions are balanced by the choice of $\beta$. Next, we will discuss the effect on the system for the two-user case, with one of each UE class (relayed or not), but the conclusions can be easily generalized for more users in both groups.

\begin{figure}[!ht]
 \centering
 \includegraphics[width=0.8\columnwidth]{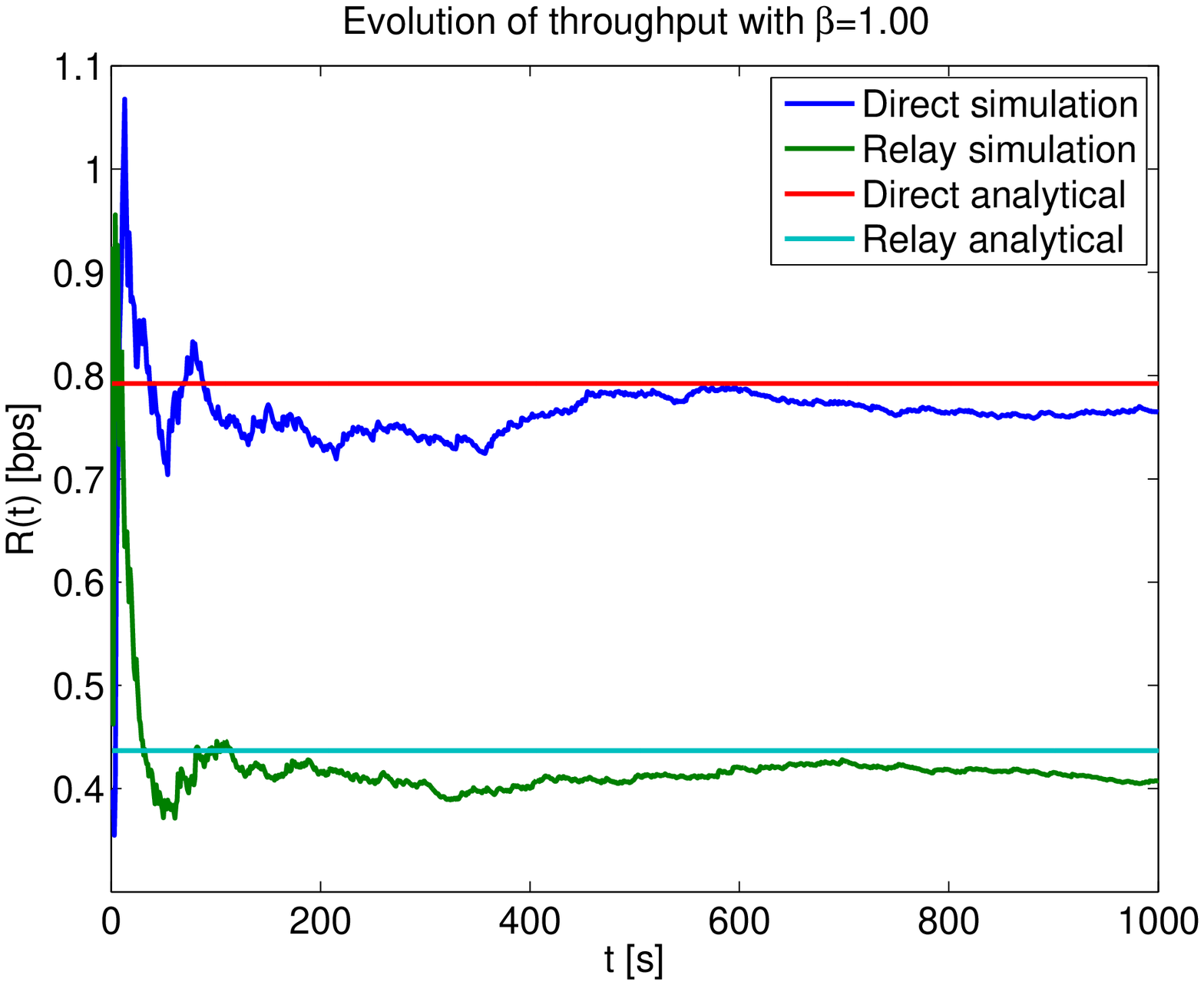}
 \caption{Evolution of the throughput of a two-user PF scheduler, with one user served through a relay.}
 \label{fig:PFtputs}
\end{figure}

Figure \ref{fig:PFtputs} shows the analytical solution for a two node scheduler and its simulation, for $\beta=1$ (no relay compensation incentives), $\alpha=0.5$ (relayed and access flows are multiplexed with a ratio of $1/2$) and $\lambda_{1,a}=\lambda_{1,r}=\lambda_2=1$ (all Rayleigh channels have unit mean). It is important to compare the rates achieved, $(0.79,0.44)\times \frac{1}{\lambda}$, with those of a two-user system without relays, $(0.75,0.75)\times \frac{1}{\lambda}.$

% If there are multiple users in each category, they achieve throughput shares that depend both on their individual channel distributions $\lambda_i$ and the relaying scheme. For example, figure \ref{fig:tputsMultiple} illustrates what happens when there are ten users (six direct and four relayed ones) and there are two possible channel values in between $\lambda_{1,a}=\lambda_{1,r}=\lambda_2=1$ or $\lambda_{1,a}=\lambda_{1,r}=\lambda_2=4$, under the same relaying circumstances above. Basically, four groups are formed (direct, good channel; direct, bad channel; relay, good channel; and relay, bad channel) and the throughputs converge to those values, which is normal given the symmetry of the ODE system with respect to equally-conditioned users.
% 
% \begin{figure}[!ht]
%  \centering
%  \includegraphics[width=0.8\columnwidth]{TputsMultiple}
%  \caption{Evolution of the throughput of a ten-user PF scheduler, with one user served through a relay (simulation only).}
%  \label{fig:tputsMultiple}
% \end{figure}

We conclude that PF schedulers without incentives are biased towards direct flows. They  hamper relayed flows heavily in exchange for small improvements in direct flows, because the former may loose their most proficient transmission opportunities while the latter only gain new transmission opportunities in suboptimal channel locations. 

\subsection{Effect of Parameter $\beta$}

\begin{figure}[!ht]
 \centering
 \includegraphics[width=0.8\columnwidth]{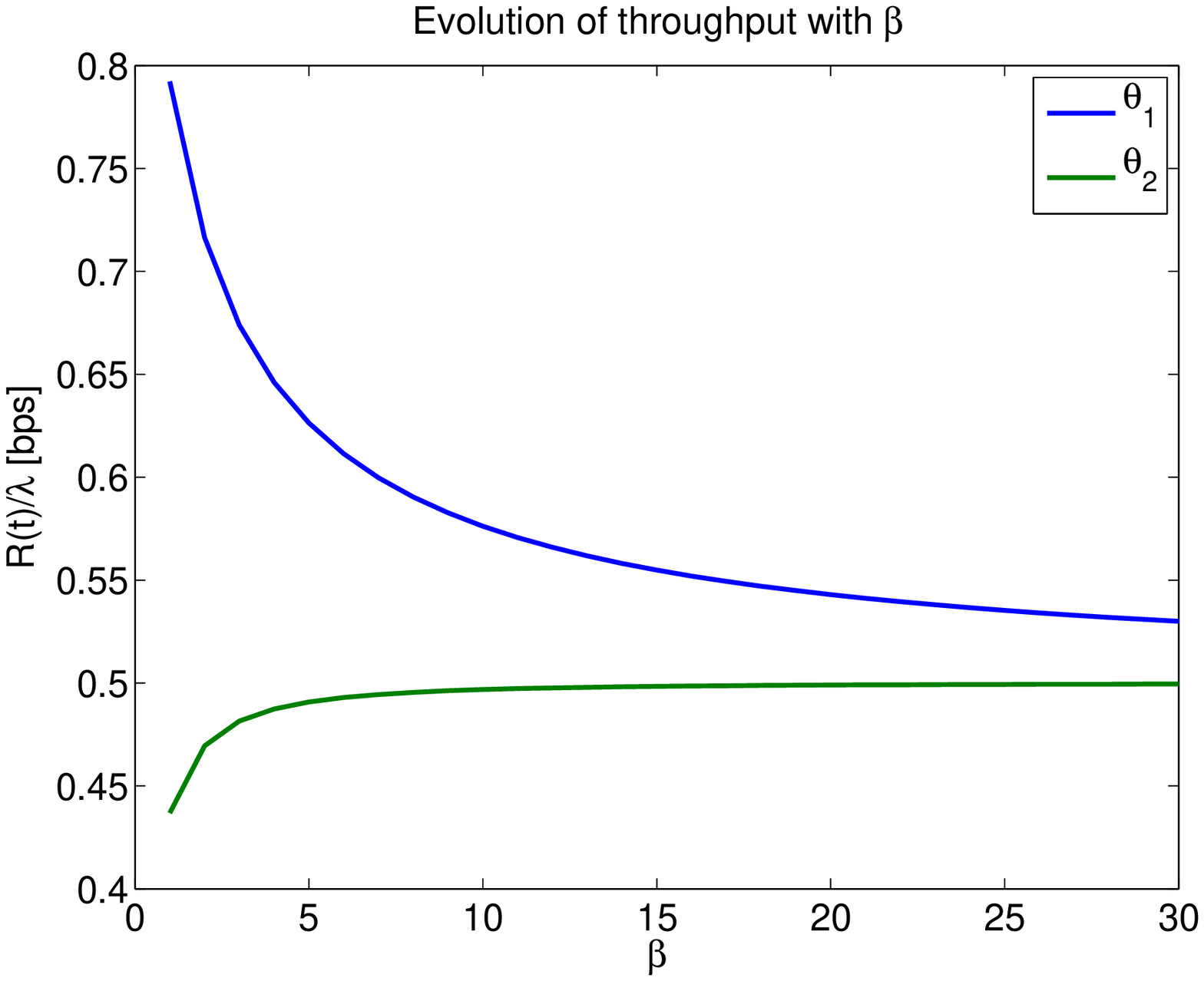}
 \caption{Evolution of the throughput of a two user PF scheduler, with one user served through a relay, as a function of $\beta$ (analytical results).}
 \label{fig:dependencybeta}
\end{figure}

We cannot correct the bias of the PF scheduler in the scenario with relays using an incentive. Figure \ref{fig:dependencybeta} shows the evolution of the two rate multipliers as a function of $\beta$, which is the incentive to increase relay flow scheduling when possible. The effect we observe is that, when we increase $\beta$, the rates converge to $(0.5,0.5)\times \frac{1}{\lambda}$ (equivalent to the RR case). Multi-user diversity gain is lost because it is increasingly likely that each user will be scheduled at fixed instants, instead of dynamically searching for the best instants. 

%This is due to the presence of only two users. However 
It is possible to generalize the conclusion to multiple users as follows: as the incentive parameter becomes increasingly high, a system with $N_D$ direct flows and $N_R$ relay flows converges to two time-multiplexed systems: one with a PF scheduler with $N_R$ users operating $\alpha$ of the time, and one with a PF scheduler with $N_D$ users operating $1-\alpha$ of the time (see Appendix \ref{app:schedulerintegral}).

\subsection{Effect of interference}

% The throughputs of the relay scenario vary with the types of the users, but they are consistently proportional to user channel gains. In figure \ref{fig:dependency} we show this relation for the two user case.

% Figure \ref{fig:dependencylambdad} shows that direct user rate $\theta_d$ is linear with $\frac{1}{\lambda_d}$ when the average relay gain $\frac{1}{\lambda_r}$ is fixed.
% 
% Figure \ref{fig:dependencylambdar} shows the opposite case for $\theta_r$.

\begin{figure}[!t]
\centering
% \subfigure[By varying $\lambda_d$]{\includegraphics[width=0.3\textwidth]{dependencylambdad}
% \label{fig:dependencylambdad}}
% \hfil
% \subfigure[By varying $\lambda_r$]{\includegraphics[width=0.3\textwidth]{dependencylambdar}
% \label{fig:dependencylambdar}}
% \hfil
% \subfigure[By varying $\frac{\lambda_{d,a}}{\lambda_{d,r}}$]{
\includegraphics[width=0.8\columnwidth]{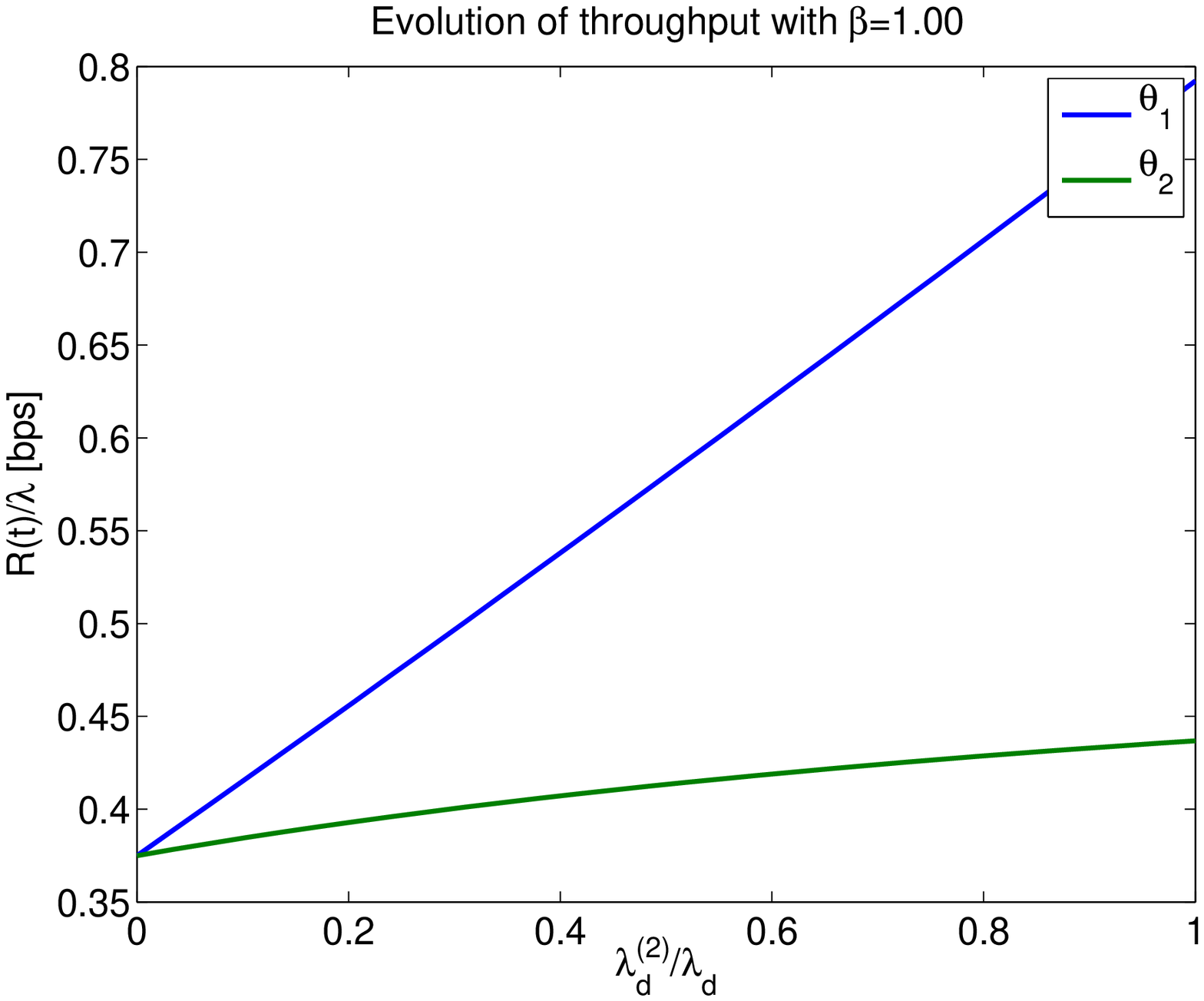}
% \label{fig:dependencyquotient}
% }
\caption{Evolution of the throughput of a two user PF scheduler, with one user served through a relay, as a function of 
% $\lambda_d$, $\lambda_r$ and 
$\frac{\lambda_{d,a}}{\lambda_{d,r}}$ (analytical results).}
\label{fig:dependency}
\end{figure}

Figure 
% \ref{fig:dependencyquotient} 
\ref{fig:dependency} 
shows the effect of higher interference, represented as an equivalent smaller channel gain, in the access phase of half-duplex relaying, due to RN transmission. On the one hand, when $\frac{\lambda_{d,a}}{\lambda_{d,r}}\rightarrow 0$, the second phase becomes less and less useful to direct users, until the scheduler converges to a part-time pure PF scheduler again ($\alpha(0.75,0.75)$). On the other hand, when $\frac{\lambda_{d,a}}{\lambda_{d,r}}= 1$, as if relay interference was completely avoided, the second phase contributes linearly to direct user throughput $\theta_d$, and due to the reduction of the direct user demand of the first phase, there is also a noticeable increment in relayed user throughput $\theta_r$. 

This means that the lack of interference management has the side effect of aggravating the conflict between scheduling and relaying.

\subsection{Effect of Half-Duplex Relaying Factor $\alpha$}
\label{sec:influencealpha}

\begin{figure}[!ht]
 \centering
 \includegraphics[width=0.8\columnwidth]{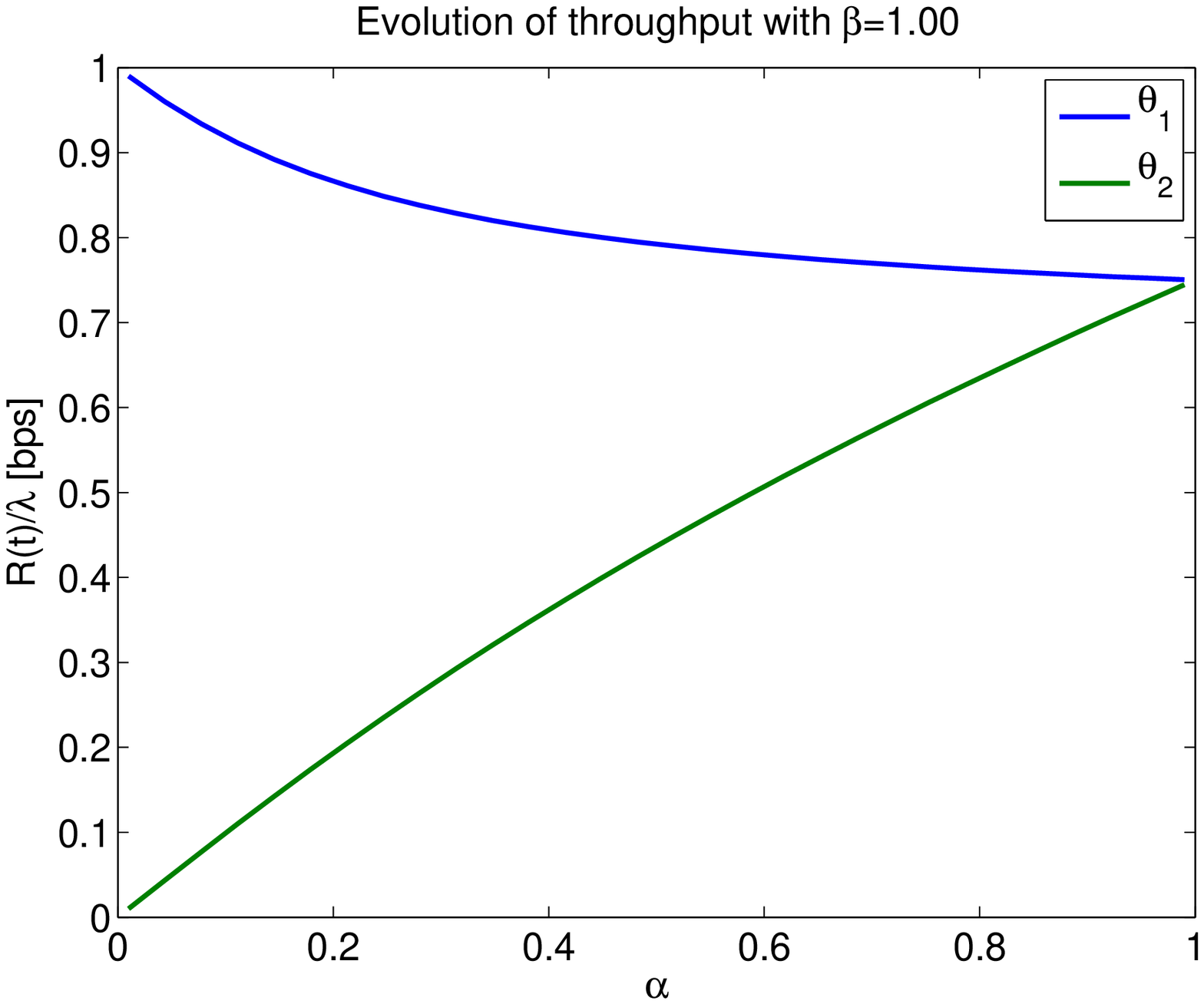}
 \caption{Evolution of the throughput of a two user PF scheduler, with one user served through a relay, as a function of $\alpha$ (analytical results).}
 \label{fig:dependencyalpha}
\end{figure}

%However, 
PF scheduling is also affected by $\alpha$, as illustrated in figure \ref{fig:dependencyalpha}. It becomes increasingly biased as $\alpha$ shrinks. In the limit when $\alpha\rightarrow1$, the PF scheduler is not biased, whereas when $\alpha\rightarrow0$ it ends up serving only direct users. 

\section{Finding an Equilibrium between RN Underutilization and Saturation}
\label{secVI}

As discussed above, relayed and access flow rates must be balanced to match inbound with outbound throughputs. If the relayed flow was higher, the relay would have to drop a significant number of packets, and if it was lower, the relay would have to transmit empty RBs. However, the LTE-A standard half-duplex timing values for TDD are quite rigid, and the optimum $\alpha$ for \eqref{eq:paralell} is not generally achievable.

Furthermore, there are two contradictory criteria in the standard. On the one hand, its \texttt{SubframeConfigurationTDD} values correspond to relays acting as receivers less than $50\%$ of the time (small $\alpha$). This implies, quite reasonably, that throughput balancing \eqref{eq:paralell} would typically rely on better radio hardware in the DeNB-RN link than in the RN-UE.

On the other hand, PF is one of the preferred schedulers in LTE, but, as shown in Section \ref{sec:influencealpha}, this family of schedulers only behaves fairly and achieves multi-user diversity gain for high values of $\alpha$. This implies that scheduling would typically be heavily hampered in the most common relaying scenario.

Therefore, since $\alpha$ values cannot be chosen to achieve complete balancing, we propose complementing the system by selecting values of $\beta$ such that, in a fraction $\alpha$ of the channel, relay flows attain the same average throughput as access flows, $\beta=\frac{\alpha\lambda_r}{(1-\alpha)\lambda_a}$. The purpose of this choice is to normalize the distribution of $\beta h_r$, for it to present an average of $\frac{\alpha}{(1-\alpha)\lambda_a}$ to the DeNB PF scheduler.

In the RN scheduler, RN access links can achieve $\overline\theta_a[t]$ independently with a different scheduler (not necessarily a PF one), although the reasons that make this scheduler more popular for the DeNB also apply to the RN.

\section{Conclusions}
\label{sec:conclusions}

Relaying will be an outstanding feature of future cellular networks, with abundant gain predicted in numerous research works. The 3GPP implementation standard for LTE-A is based on Layer-3 data forwarding and fixed parametric timing of the two relaying phases. We have shown that this rigid relaying implementation generates conflicts with other aspects of the network, such as scheduling, raising problems that have been ignored by the flexible analytical models in the research literature.

The standard defines fixed partitioning between relay and access links, creating an imbalance whereby part of their capacities may be underutilized. We propose interference management to concentrate transmission on a fraction of the total resources available to the underutilized link. This may mitigate the undesired interference increase due to relaying. Our analysis of a relay system with PF schedulers shows that interference with direct users during the access phase is also harmful for scheduling. There is a trade-off, because reducing interference subframes reduces RN-UE transmission opportunities, constraining scheduling flexibility and, thus, scheduling gains.

Schedulers may also experience other negative effects. The worst would be a loss of multi-user diversity due to half-duplex relay operation, leading to suboptimal scheduling. We show that, even though schedulers can be modified to compensate for link unbalancing by introducing incentives, the cost is an even higher loss of multi-user diversity, to the point that the PF scheduler may achieve RR performance at most. We propose seeking an intermediate point with incentives designed to balance DeNB-RN and RN-UE links. 

The set of relay configuration parameters in the standard is reasonable and extensive, but it seems to ignore some scenarios that may appear in a real deployment. Particularly, when potential multi-user diversity gain is high (for instance, with many relays and UEs), we have shown that it might be beneficial to grant more resources to relay channels than what currently allowed. 

% RN location is a critical performance parameter. We have shown that it is possible to develop admission control strategies for randomly located RNs in which some of them may be discarded when they are unlikely to contribute to a better network capacity.

\appendix 
% \onecolumn
\section{Average Throughput in PF Schedulers}
\label{app:schedulerintegral}

% \subsection{Single-Cell Scheduling}
In \cite{Kushner2004} a set of users $i\in[1,n]$ experiencing Rayleigh fading is considered, with exponentially distributed instantaneous SNR $h_i\sim \textnormal{Exp}(\lambda_i)$. The achievable rate per resource use is $\theta_i[t]\simeq h_i$, and the scheduling function is $s[t]\in[1,n] = \arg_i\max \frac{\theta_i[t]}{\overline\theta_i}$. In the permanent state of the ODE the average rate of each user will converge to the constant value: 
\begin{equation}
\lim_{T\rightarrow \infty}\frac{1}{T}\sum_{t=1,\{t:s[t]=i\}}^T(\theta_i[t])\rightarrow \overline\theta_i
\end{equation}

% The conditioned averages to complete the ODE 
% assuming these averages remain constant, as follows:
% \begin{equation}
% \label{eq:avgint}
% \begin{split}
%   \Ex{h_i[t]/\overline{h}_i>h_j[t]/\overline{h}_j\forall j}{h_i[t]}&=\int_0^\infty x f_{h_i[t]}(x)\prod_{j\neq i}  F_{h_j[t]}\left(x\frac{\overline{h}_j}{\overline{h}_i}\right)dx\\ 
%    &=\int_0^\infty x \lambda_ie^{-\lambda_i x}\prod_{j\neq i} 1-e^{-\lambda_j x \frac{\overline{h}_j}{\overline{h}_i}}dx\\
% %    &=\int_0^\infty x \lambda_i e^{-x \sum_{j} \lambda_j \frac{\overline{h}_j}{\overline{h}_i}}dx\\
% %    &=\frac{\lambda_i\overline{h}_i^2}{\left( \sum_{j} \lambda_j\overline{h}_j\right)^2}\\
%   \end{split}
% \end{equation}
% 
% For a small number of users, we can extend the product completely, but this requires solving a different integral for each case
The ODE for the two-user case is:
% :
% \begin{equation}
% \begin{split}
%    &=\int_0^\infty x \lambda_1 \left[e^{-x\lambda_1}-e^{\lambda_1+\lambda_2 \frac{\overline{h}_1}{\overline{h}_2}}\right]dx\\
%    &=\frac{1}{\lambda_1}-\frac{\lambda_1\overline{h}_1^2}{\left( \lambda_1\overline{h}_1+\lambda_2\overline{h}_2\right)^2}\\
%   \end{split}
% \end{equation}
% 
% Taking this to the
%producing the ODE
\begin{equation}
 \begin{array}{ccccc}
 0&=&\frac{1}{\lambda_1}-\frac{\lambda_1\overline{\theta}_1^2}{\left( \lambda_1\overline{\theta}_1+\lambda_2\overline{\theta}_2\right)^2}-\overline{\theta}_1\\
 0&=&\frac{1}{\lambda_2}-\frac{\lambda_2\overline{\theta}_2^2}{\left( \lambda_1\overline{\theta}_1+\lambda_2\overline{\theta}_2\right)^2}-\overline{\theta}_2\\
 \end{array}
\end{equation}
% Other interesting simplifications exist, such as considering all channels equal $\lambda_j=\lambda \forall j$
% And therefore, using the condition that the solution is stable (derivatives become zero) the ODE can be rewritten as
% \begin{equation}
%  \begin{array}{ccccc}
%  \overline{\theta}_1^3&=&\frac{\lambda_1}{\left( \sum_{j} \lambda_j \frac{1}{\overline{\theta}_j}\right)^2}\\
%   \overline{\theta}_2^3&=&\frac{\lambda_2}{\left( \sum_{j} \lambda_j \frac{1}{\overline{\theta}_j}\right)^2}\\
%   \vdots&&\vdots&&\vdots\\
%   \overline{\theta}_n^3&=&\frac{\lambda_n}{\left( \sum_{j} \lambda_j \frac{1}{\overline{\theta}_j}\right)^2}\\
%  \end{array}
% \end{equation}
A more general solution for $n$ users is also given in \cite{Kushner2004}. 

We provide solve \eqref{eq:ode} for the relayed case, which is similar to that in \cite{Kushner2004}, with the following modifications:
% \subsection{Cell and Relay Scheduling}
% A solution of \ref{eq:ode} for the scheduling of the first hop of a DeNB, when there are $n_d$ direct flows and $n_r$ flows towards the relay. 
We consider a PF scheduler operating independently over $\tau_r+\tau_a$ RBs. For $\tau_r$ of these, direct and relay flows may be scheduled, while only direct flows can be scheduled for the remaining $\tau_a$. An incentive parameter $b_i$ is employed to tune the scheduler to enforce fairness. In addition, during $\tau_r$ and $\tau_a$, direct users may experience different channels, modeled with two independent variables with averages $\lambda_{i,r}$ and $\lambda_{i,a}$, due to the activation of the RN in the access phase. 

Expression \eqref{eq:oderelay} is the resulting ODE. The averages in the three terms can be computed analogously to as in \cite{Kushner2004} with the trivial inclusion of $b_i$ values in the scheduling part and differentiating $\lambda_a$ and $\lambda_r$ values according to their definition.

For the two-user case,
% , we get
% \begin{equation}
% \begin{split}
%    =&\alpha \int_0^\infty x \lambda_{1,r} \left[e^{-x\lambda_{1,r}}-e^{\lambda_{1,r}+\lambda_{2,r} \frac{\overline{h}_2b_1}{\overline{h}_1b_2}}\right]dx \\
%     &+(1-\alpha)\int_0^\infty x \lambda_{1,a} \left[e^{-x\lambda_{1,r}}-e^{\lambda_{1,a}+\lambda_{2,a} \frac{\overline{h}_2b_1}{\overline{h}_1b_2}}\right]dx\\
%    =&\alpha\left[\frac{1}{\lambda_{1,r}}-\frac{(\overline{h}_1b_1)^2\lambda_{1,r}}{\left(\lambda_{1,r}\overline{h}_1+\lambda_{2,r}\overline{h}_2\right)^2}\right]\\
%     &+(1-\alpha)\left[\frac{1}{\lambda_{1,a}}-\frac{(\overline{h}_1b_1)^2\lambda_{1,a}}{\left( \lambda_{1,a}\overline{h}_1+\lambda_{2,a}\overline{h}_2\right)^2}\right]\\
%   \end{split}
% \end{equation}
% 
% If 
assuming the first user is direct and the second is relayed, the exact solution is:
% $b_1=1$, $b_2=\beta$, and $\lambda_{2,a}=0$. Substituting in the ODE
\begin{equation}
 \begin{array}{ccccc}
 0&=&\alpha\left[\frac{1}{\lambda_{1,r}}-\frac{\lambda_{1,r}\overline{\theta}_1^2}{\left( \lambda_{1,r}\overline{\theta}_1+\lambda_{2,r}\beta\overline{\theta}_2\right)^2}\right]\\
 &&\quad+(1-\alpha)\left[\frac{1}{\lambda_{1,a}}-\overline{\theta}_1\right]-\overline{\theta}_1\\
 0&=&\alpha\left[\frac{1}{\lambda_{2,r}}-\frac{(\overline{\theta}_2\beta)^2\lambda_{2,r}}{\left( \lambda_{1,r}\overline{\theta}_1+\lambda_{2,r}\beta\overline{\theta}_2\right)^2}\right]-\overline{\theta}_2\\
 \end{array}
\end{equation}

Due to the additive term $+(1-\alpha)\left[\frac{1}{\lambda_{1,a}}-\overline{\theta}_1\right]$, this system has a solution that is not a trivial transformation of the original one in \cite{Kushner2004}. 
% Therefore, we have solved this system numerically, as shown in figure \ref{fig:tputs}.

% \begin{equation}
%  \begin{array}{ccccc}
%  \overline{\theta}_1^3&=&\alpha\frac{\lambda_{1,r}}{\left( \sum_{j} \lambda_{j,r} \frac{b_j}{\overline{\theta}_j}\right)^2}&+&(1-\alpha)\frac{\lambda_{1,a}}{\left( \sum_{j} \lambda_{j,a} \frac{b_j}{\overline{\theta}_j}\right)^2}\\
%   \overline{\theta}_2^3&=&\alpha\frac{\lambda_{2,r}}{\left( \sum_{j} \lambda_{j,r} \frac{b_j}{\overline{\theta}_j}\right)^2}&+&(1-\alpha)\frac{\lambda_{2,a}}{\left( \sum_{j} \lambda_{j,a} \frac{b_j}{\overline{\theta}_j}\right)^2}\\
%   \vdots&&\vdots&&\vdots\\
%   \overline{\theta}_{n_d}^3&=&\alpha\frac{\lambda_{n_d,r}}{\left( \sum_{j} \lambda_{j,r} \frac{b_j}{\overline{\theta}_j}\right)^2}&+&(1-\alpha)\frac{\lambda_{n_d,a}}{\left( \sum_{j} \lambda_{j,a} \frac{b_j}{\overline{\theta}_j}\right)^2}\\
%  \overline{\theta}_{n_d+1}^3&=&\alpha\frac{\lambda_{n_d+1}}{\left( \sum_{j} \lambda_{j,r} \frac{b_j}{\overline{\theta}_j}\right)^2}\\
%   \overline{\theta}_{n_d+2}^3&=&\alpha\frac{\lambda_{n_d+2}}{\left( \sum_{j} \lambda_{j,r} \frac{b_j}{\overline{\theta}_j}\right)^2}\\
%   \vdots&&\vdots\\
%   \overline{\theta}_{n_d+n_r}^3&=&\alpha\frac{\lambda_{n_d+n_r}}{\left( \sum_{j} \lambda_{j,r} \frac{b_j}{\overline{\theta}_j}\right)^2}\\
%  \end{array}
% \end{equation}

%Moreover,
% , as $b_i$ equals $1$ for direct users and $\beta$ for relay users, 
When the incentive becomes too large ($\beta\rightarrow \infty$), the ODE system converges to:
\begin{equation}
% \lim_{\beta\rightarrow \infty}
% \left\{
% %  \begin{array}{ccccc}
% %  0&=&\alpha\left[\frac{1}{\lambda_{1,r}}-\frac{\lambda_{1,r}\overline{\theta}_1^2}{\left( \lambda_{1,r}\overline{\theta}_1+\lambda_{2,r}\beta\overline{\theta}_2\right)^2}\right]+(1-\alpha)\left[\frac{1}{\lambda_{1,a}}-\overline{\theta}_1\right]-\overline{\theta}_1\\
% %  0&=&\alpha\left[\frac{1}{\lambda_{2,r}}-\frac{(\overline{\theta}_2\beta)^2\lambda_{2,r}}{\left( \lambda_{1,r}\overline{\theta}_1+\lambda_{2,r}\beta\overline{\theta}_2\right)^2}\right]-\overline{\theta}_2\\
% %  \end{array}
%  \right\}
%  =
% \left\{
 \begin{array}{ccccc}
 0&=&(1-\alpha)\left[\frac{1}{\lambda_{1,a}}\right]-\overline{\theta}_1\\
 0&=&\alpha\left[\frac{1}{\lambda_{2,r}}\right]-\overline{\theta}_2\\
 \end{array}
\end{equation}

\section*{Acknowledgements}
This research has been supported by grants CALM (TEC2010-21405-C02-01) and FPU  2012/01319, Mineco, Spain; AtlantTIC (CN 2012/260), European Regional Development Fund (ERDF) and Xunta de Galicia, Spain; and MEFISTO (10TIC006CT), Xunta de Galicia, Spain.

\bibliographystyle{IEEEtran}
\bibliography{LTE,Schedulers}

\begin{thebibliography}{10}
\providecommand{\url}[1]{#1}
\csname url@samestyle\endcsname
\providecommand{\newblock}{\relax}
\providecommand{\bibinfo}[2]{#2}
\providecommand{\BIBentrySTDinterwordspacing}{\spaceskip=0pt\relax}
\providecommand{\BIBentryALTinterwordstretchfactor}{4}
\providecommand{\BIBentryALTinterwordspacing}{\spaceskip=\fontdimen2\font plus
\BIBentryALTinterwordstretchfactor\fontdimen3\font minus
  \fontdimen4\font\relax}
\providecommand{\BIBforeignlanguage}[2]{{%
\expandafter\ifx\csname l@#1\endcsname\relax
\typeout{** WARNING: IEEEtran.bst: No hyphenation pattern has been}%
\typeout{** loaded for the language `#1'. Using the pattern for}%
\typeout{** the default language instead.}%
\else
\language=\csname l@#1\endcsname
\fi
#2}}
\providecommand{\BIBdecl}{\relax}
\BIBdecl

\bibitem{mogensen2009lte}
P.~E. Mogensen, T.~Koivisto, K.~I. Pedersen, I.~Z. Kov\'{a}cs, B.~Raaf,
  K.~Pajukoski, and M.~J. Rinne, ``{LTE-advanced: the path towards gigabit/s in
  wireless mobile communications},'' in \emph{1st International Conference on
  Wireless Communication, Vehicular Technology, Information Theory and
  Aerospace \& Electronic Systems Technology}.\hskip 1em plus 0.5em minus
  0.4em\relax IEEE, 2009, pp. 147--151.

\bibitem{journals/cm/LoaWSYCHX10}
K.~Loa, C.-C. Wu, S.-t. Sheu, Y.~Yuan, M.~Chion, D.~Huo, and L.~Xu,
  ``{IMT-advanced relay standards [WiMAX/LTE Update].}'' \emph{IEEE
  Communications Magazine}, vol.~48, no.~8, pp. 40--48, 2010.

\bibitem{LTE36216PHYrelay}
3GPP, ``{Universal Mobile Telecommunications System (UMTS); Evolved Universal
  Terrestrial Radio Access (E-UTRA); Physical layer for relaying operation},''
  \emph{ETSI TS 136 216}, no. 10.3.1, pp. 1--18, 2011.

\bibitem{Specification2012c}
------, ``{LTE; Evolved Universal Terrestrial Radio Access (E-UTRA); Long Term
  Evolution (LTE) physical layer; General description (3GPP TS 36.201 version
  8.3.0 Release 8)},'' \emph{ETSI TR 136 201}, no. 8.3.0, pp. 1--15, 2012.

\bibitem{Karaer2009optimization}
A.~Karaer, O.~Bulakci, S.~Redana, and J.~H\"{a}m\"{a}l\"{a}inen, ``{Uplink
  performance optimization in relay enhanced LTE-Advanced networks},'' in
  \emph{IEEE 20th International Symposium on Personal, Indoor and Mobile Radio
  Communications}, 2009, pp. 360--364.

\bibitem{conf/vtc/SalehRHR09}
A.~B. Saleh, S.~Redana, B.~Raaf, and J.~H\"{a}m\"{a}l\"{a}inen, ``{Comparison
  of Relay and Pico eNB Deployments in LTE-Advanced.}'' in \emph{IEEE VTC
  Fall}, 2009.

\bibitem{conf/vtc/SalehRHRRW09Performance}
A.~B. Saleh, B.~Raaf, S.~Redana, T.~Riihonen, J.~H\"{a}m\"{a}l\"{a}inen, and
  R.~Wichman, ``{Performance of Amplify and Forward and Decode and Forward
  Relays in LTE-Advanced.}'' in \emph{IEEE VTC Fall}, 2009.

\bibitem{journals/jece/SalehRHR10}
A.~B. Saleh, S.~Redana, J.~H\"{a}m\"{a}l\"{a}inen, and B.~Raaf, ``{On the
  coverage extension and capacity enhancement of inband relay deployments in
  LTE-Advanced networks},'' \emph{Journal of Electrical and Computer
  Engineering}, 2010.

\bibitem{vienaVTC2010}
J.~C. Ikuno, M.~Wrulich, and M.~Rupp, ``{System level simulation of LTE
  networks},'' in \emph{Proc. IEEE 71st Vehicular Technology Conference},
  Taipei, Taiwan, May 2010.

\bibitem{LTE36213PHYproc}
3GPP, ``{LTE; Evolved Universal Terrestrial Radio Access (E-UTRA); Physical
  layer procedures (3GPP TS 36.213 version 8.8.0 Release 8)},'' \emph{ETSI TS
  136 213}, no. 8.8.8, pp. 1--79, 2009.

\bibitem{conf/globecom/HuangUAHB10}
X.~Huang, F.~Ulupinar, P.~Agashe, D.~Ho, and G.~Bao, ``{LTE Relay Architecture
  and Its Upper Layer Solutions.}'' in \emph{IEEE GLOBECOM}, 2010.

\bibitem{LTE36300UTRANgen}
3GPP, ``{LTE; Evolved Universal Terrestrial Radio Access (E-UTRA) and Evolved
  Universal Terrestrial Radio Access Network (E-UTRAN); Overall description;
  Stage 2 (3GPP TS 36.300 version 10.7.0 Release 10)},'' \emph{ETSI TS 136
  300}, no. 10.7.0, pp. 1--204, 2012.

\bibitem{journals/ejwcn/DopplerRWRW09}
K.~Doppler, S.~Redana, M.~W\'{o}dczak, P.~Rost, and R.~Wichman, ``{Dynamic
  resource assignment and cooperative relaying in cellular networks: Concept
  and performance assessment},'' \emph{EURASIP Journal on Wireless
  Communications and Networking}, 2009.

\bibitem{conf/icc/RongEH10}
L.~Rong, S.~E. Elayoubi, and O.~B. Haddada, ``{Impact of Relays on LTE-Advanced
  Performance.}'' in \emph{IEEE International Conference on Communications},
  2010.

\bibitem{136211LTEPHYr10}
3GPP, ``{LTE; Evolved Universal Terrestrial Radio Access (E-UTRA); Physical
  channels and modulation (3GPP TS 36.211 version 10.4.0 Release 10)},''
  \emph{ETSI TR 136 211}, no. 10.4.0, pp. 1--103, 2012.

\bibitem{Kushner2004}
H.~J. Kushner and P.~A. Whiting, ``{Convergence of Proportional-Fair Sharing
  Algorithms Under General Conditions},'' \emph{IEEE Transactions on Wireless
  Communications}, vol.~3, no.~4, pp. 1250--1259, Jul. 2004.

\bibitem{Gitlin}
H.-Y. Wei and R.~Gitlin, ``{Incentive scheduling for cooperative relay in
  WWAN/WLAN two-hop-relay network},'' \emph{IEEE Wireless Communications and
  Networking Conference}, vol.~3, pp. 1696--1701, 2005.

\end{thebibliography}

\end{document}